\documentclass[prl,floatfix, twocolumn]{revtex4}
\usepackage{amssymb}

\usepackage{graphicx}

\begin{document}

\title{Theory of cooling by flow through narrow pores}

\author{William J. Mullin}
\email{mullin@physics.umass.edu}
\affiliation{Physics Department, Hasbrouck Laboratory, University of
Massachusetts, Amherst, MA 01003} 
\author {Neal Kalechofsky}
\email{kalechofsky@ma.oxinst.com}
\affiliation{Oxford Instruments
America Inc., 130A Baker Avenue, Concord, MA, 01742, USA}

\date{\today }

\begin{abstract}
We consider the possibility of adding a stage to a dilution refrigerator to
provide additional cooling by ``filtering out'' hot atoms. Three methods are
considered: 1) Effusion, where holes having diameters larger than a
mean-free path allow atoms to pass through easily; 2) Particle
waveguide-like motion using very narrow channels that greatly restrict the
quantum states of the atoms in a channel. 3) Wall-limited diffusion through
channels, in which the wall scattering is disordered so that local density
equilibrium is established in a channel. We assume that channel dimension
are smaller than the mean-free path for atom-atom interactions. The particle
waveguide and the wall-limited diffusion methods using channels on order of
the de Broglie wavelength give cooling. Recent advances in nano-filters give
this method some hope of being practical.
\end{abstract}


\maketitle


We investigate here the possibility cooling a gas by ``filtering out'' hot
atoms, perhaps as an adjunct to a dilution refrigerator with solutions of $%
^{3}$He in liquid $^{4}$He.\cite{LT24} 
Our method involves passing a
degenerate Fermi gas through narrow constrictions formed by pores in a
partition. Under certain conditions we find that this particle ``leakage''
allows a lowering of the temperature of the remaining gas. Our study adds to
many recent experimental and theoretical analyses of quantum size effects in
the behavior of particles in nanopores (as in Refs.
\cite{Nanophysics,Pickett,Cole}).

An initial idea of how to remove the hot atoms is suggested by a technique
used in electron heterojunction physics,\cite{Hetero,Szafer,Butt,Bag} where
the electron gas is passed through a narrow constriction formed by a gate
potential. Because the constriction is narrow the bands of states allowed in
this ``particle waveguide'' are widely separated, which means that not all
energies are allowed through. Adjusting the states in the channel can allow
selective passage particles in states at the Fermi energy, so that one
removes only hot atoms. In such an approach the constriction must be of
order of the de Broglie wavelength of atoms at the top of the Fermi surface,
which is roughly the separation between fermions. Because of the limited
states (that is, bands) we characterize channels of this size as ``\emph{%
narrow}.'' We will see that this approach can be made to work under
appropriate conditions on the nano-pores. 

An alternative possibility involves channels or pores with diameters much
larger than the de Broglie wavelength but still less than the mean-free path
of the fermions in the gas, which can be large due to its $1/T^{2}$
behavior. We call these ``\emph{wide}'' pores. 

We will consider different kinds of flow through the holes of the two sizes
mentioned: \emph{effusion,} \emph{waveguide flow,} and \emph{wall-limited
diffusion}. In effusion, the holes are by definition wide and the states in
the channel remain three-dimensional with no banding. Moreover the walls are
sufficienly smooth that the particles undergo no back scattering nor do they
come into equilibrium with the channel walls. Effectively all that the pores
do is to allow particles already directed in the positive $z$-direction to
pass through the membrane. We might hope that this would allow cooling
because the intensity of fast atoms passing through a hole is larger than
that of the slow atoms and these, on average, carry more energy. Indeed it
is a standard textbook exercise\cite{Pathria} to show that this works 
for a Boltzmann gas. But does it work for a degenerate Fermi gas?

In waveguide flow the channels are narrow enough to have well-defined bands.
Again the particles are assumed to undergo no back scattering nor do they
come into thermal equilibrium with the channel walls. Nevertheless because
of the bands only particles with certain energies are allowed in the
channel. If, for example, the Fermi energy of the gas in the container is
coincident with the bottom of the lowest band one might expect that only
high energy particles would get through the holes and the remaining gas
would be cooled.

A third situation is wall-limited diffusion or Knudsen flow,\emph{\ }which
could occur in wide or narrow pores. There is scattering at the walls
including back scattering. The rate of diffusion in the channel depends on
the diameter of the pore, the particle velocity, and the density gradient
that maintains the flow. In this and the above methods we will assume the
density difference is maintained by pumping away the particles that pass
through the membrane. Both effusion and wall-limited diffusion were
considered by one of the authors some time ago for enhancing polarization.%
\cite{Mull}

If the Fermi gas in the cooling cell is, say, a 1\% solution of $^{3}$He in
liquid $^{4}$He at millikelvin temperature, then the Fermi temperature is
124 mK, the de Broglie wavelength about 3 nm, and the mean-free path is
about 1 $\mu $m at $T=15$ mK.

Suppose the number of atoms in a box (B1) is $N$ with density $n.$ The atoms
pass through a membrane and enter a second box (B2) maintained (by pumping)
at a much smaller particle number $N_{2}$ and density $n_{2}$. The membrane
contains a great number $M$ of holes each of diameter $d$ and of total hole
area $A$. (See Fig.\ 1.) We will compute particle current $J_{N}$ and energy current $J_{E}$
passing through the holes$.$ The rates of particle and energy change, $%
dN/dt=AJ_{N}$ and $dE/dt=AJ_{E}$ in the container B1 combine in 
\begin{equation}
\frac{dE}{dt}=C_{V}\frac{dT}{dt}+\left( \frac{\partial E}{\partial N}\right)
_{T,V}\frac{dN}{dt}  \label{dedt}
\end{equation}
to give the cooling power $C_{V}dT/dt$ where $C_{V}$ is the heat capacity at
constant volume in B1.
\begin{figure}[h]
\centering
\includegraphics[width=3.0in]{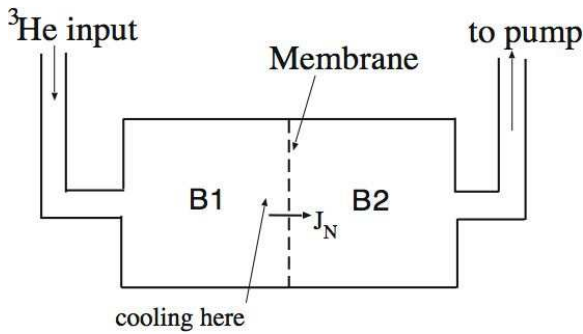}
\caption{Schematic diagram of the apparatus.  $^{3}$He in solution
with liquid $^{4}$He enters the cooling chamber B1, passes
selectively through the membrane into B2.  A gas at lowered
temperature remains in B1.  The $^{3}$He density in B2 is kept low by
pumping.  The gas is recycled, after being recooled by a dilution
fridge, back into B1.}
\label{apparatus}
\end{figure}

In the case of effusion the particle current is 
\begin{equation}
J_{N}(1{\ \rightarrow }2)=\frac{2}{h^{3}}\int_{-\infty }^{+\infty
}dp_{x}\int_{-\infty }^{+\infty }dp_{y}\int_{0}^{+\infty }dp_{z}\frac{p_{z}}{%
m}n_{p}  \label{Eq1}
\end{equation}
where $n_{p}$ is the momentum distribution function and the factor of 2
accounts for spin degeneracy. There is a similar expression for $J_{E}.$ For
a classical gas passing through wide holes we find a cooling power given by $%
-nk_{B}AT\left( k_{B}T/2\pi m\right) ^{1/2}.$ The minus sign implies cooling
in agreement with the usual textbook treatment.\cite{Pathria} In the
degenerate limit effusive cooling fails.

For particle waveguide motion (in narrow holes) we need to take into account
the banding of the states due to the limited transverse motion; the energies
are given by $\epsilon _{nz}=\epsilon _{n}+\epsilon _{z}$ with $\epsilon
_{z}=\hbar ^{2}k_{z}^{2}/2m.$ We will assume a square cross section with
width $d$ so that $\epsilon _{n}=\epsilon _{0}\eta _{n}$ where $\epsilon
_{0}=\pi ^{2}\hbar ^{2}/(md^{2})$ and $\eta
_{n}=1,2.5,...(n_{x}^{2}+n_{y}^{2})/2,....$ (Using a circular cross section
gives no significant differences in our final results.) In this case Eq.\ (%
\ref{Eq1}) is replaced by 
\begin{equation}
J_{N}(1{\rightarrow }2)=\frac{2}{d^{2}h}\sum_{n}\int_{0}^{+\infty }dp_{z}%
\frac{p_{z}}{m}\frac{1}{(e^{\beta (\epsilon _{z}+\epsilon _{n}-\mu )}+1)}
\end{equation}
with $\mu $ the chemical potential ($\approx \epsilon _{F}$ at low $T).$ In
terms of the one-dimensional Fermi integrals $F_{l}(\gamma )=\int_{0}^{%
\infty }dzz^{l}1/(e^{z-\gamma }+1)$, the cooling power for the waveguide
case is 
\begin{equation}
C_{V}\frac{dT}{dt}=-\frac{2A}{d^{2}h\beta ^{2}}p_{W}  \label{GenForm}
\end{equation}
where 
\begin{equation}
p_{W}\equiv \sum_{n=1}^{\infty }\left[ F_{1}(\alpha _{n})-\left( \xi -\beta
\epsilon _{n}\right) \ln (1+e^{\alpha _{n}})\right] .  \label{main}
\end{equation}
The filling of the bands in the channels is determined by $\alpha _{n}=\beta
(\mu -\epsilon _{n})$ with $\beta =(k_{B}T)^{-1}$and $\xi =\beta \left( 
\partial E/\partial N\right) _{T,V}$ is a property of the gas in the cooling
chamber (B1). For cooling $p_{W}$ must be positve. If $\gamma =\beta \mu $
and $t=T/T_{F}$ then the relation between density and chemical potential
leads to the usual transcendental equation for $\mu :$%
\begin{equation}
\frac{3}{2}t^{3/2}F_{1/2}(\gamma )=1  \label{selfconsis}
\end{equation}
Further, taking the derivative of $E$ versus $N$ gives 
\begin{equation}
\xi =\frac{G_{3/2}(\gamma )}{G_{1/2}(\gamma )}  \label{x}
\end{equation}
where $G_{l}(\gamma )=\int_{0}^{\infty }dzz^{l}e^{z-\gamma }/(e^{z-\gamma
}+1)^{2}=lF_{l-1}.$ Given a value of $t$ from the temperature and the
concentration, we must solve Eq.~(\ref{selfconsis}) for $\gamma ;$ we put
that into Eq.~(\ref{x}) to get $\xi .$ We find it convenient to introduce
the the ratio of lowest band edge to Fermi energy $y\equiv \epsilon
_{0}/\epsilon _{F}$ so that $\beta \epsilon _{n}=y\eta _{n}/t.$

For numerical calculations we \emph{fix} $T$ at 15 mK as a reasonable
incoming temperature in the cooling cycle. Then, if the number of holes in
the membrane is $M$ =10$^{11},$ the prefactor in Eq.~(\ref{GenForm}) is 13 $%
\mu $W. In the case of wide holes (where waveguide flow becomes effusion),
the sum over bands in Eq.\ (\ref{GenForm}) becomes a double integral over
transverse momenta and in this continuum limit we find 
\begin{equation}
p_{W}\approx \frac{\pi }{2}\frac{t}{y}\left[ F_{2}(\gamma )-\xi F_{1}(\gamma
)\right] \mathrm{~~continuum~limit}  \label{contlim}
\end{equation}
We have made no assumptions here about whether the system is degenerate or
not---just that the channel states are now continuous. If the factor in
square brackets is positive, then it says that the smaller $y$, the larger
is the cooling. This result stems from just having bigger and bigger holes,
allowing more hot atoms out. But of course there is a limit to how big the
holes can be to maintain a pressure differential across the membrane.

Next consider the highly degenerate limit of the last form, for which $%
F_{n}=\gamma ^{n}/n$ and $\xi =\gamma ,$ so that 
\begin{equation}
p_{W}\approx -\frac{\pi }{12}\frac{t}{y}\gamma ^{3}\mathrm{%
~~degenerate~continuum~limit}  \label{degcont}
\end{equation}
which agrees with a direct calculation---no cooling occurs in that case,
because there are many channel states below the Fermi energy allowing low
energy atoms to escape. Indeed the gas left behind can then end with a
higher temperature.

For the fully classical limit we take $e^{\gamma }$ to be very large so that 
\begin{equation}
p_{W}\approx \frac{\sqrt{\pi }}{3}\frac{1}{t^{1/2}y}\text{~~classical
continuum limit.}  \label{classLim}
\end{equation}
This result again agrees with the textbook classical effusion result.

In Fig.~2 we show numerical results based on Eq.~(\ref{main}). It is useful
to plot $yp_{W}$ rather than $p_{W}$ because then all the plots collapse
onto a single curve for $t>1$, as implied by Eq.~(\ref{contlim}). The $t$
dependence for large $t$ is $t^{-1/2}$ and we are indeed in the asymptotic
region described by Eq.~(\ref{classLim}). We see that $y=1$ is a special
point with the lowest band edge lined up with the Fermi energy. (Recall
however, that the cooling power prefactor, which is not included in $p_{W}$
contains a $T^{2}$ so all cooling powers go to zero with decreasing $T$.)
When the lowest band edge is lower than the Fermi energy ($y<1)$ we lose
cooling at very low $T$ (as in Eq.\ (\ref{degcont})), but surprisingly we 
\emph{regain} it for larger $T$ due to the contributions of the higher bands 
$.$ Finally for all band edges above the Fermi energy ($y>1)$ we get cooling
for all $T$, but, because the curves in the figure are multiplied by the
factor $y,$ the actual cooling power is diminished when we divide out this
factor, and these cases are not as useful. 
\begin{figure}[h]
\centering
\includegraphics[width=3.0in]{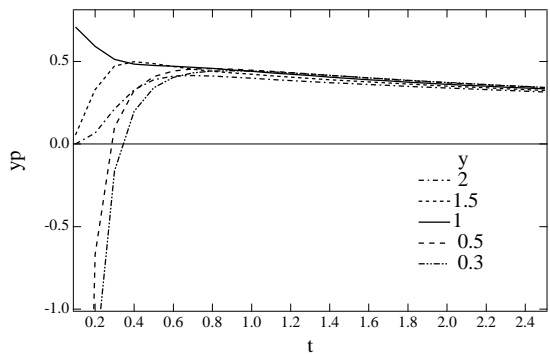}
\caption{Reduced waveguide cooling power as a function of temperature
parameter $t$ for various band-edge $y$ values for a square cross
section.  Because of the multiplicative factor $y$ all curves collapse
onto the same classical limit curve at large $t$.}
\label{square}
\end{figure}

The parameters we have used are not independent. Because $T_{F}$ depends on $%
x^{2/3}$, where $x$ is the concentration, and $\epsilon _{0}$ depends
on $d^{-2},$ then for $T=15$ mK we find the relation
\begin{equation}
t=1.8\times 10^{-2}y\left( d\mathrm{(nm)}\right) ^{2}  \label{tvsyd}
\end{equation}
From the plot, we see that if we have $y<1$ we get a $1/y$ enhancement in
cooling power, but to gain that we must also have $t\gtrsim 0.4$ to avoid
heating. For $y>1$ we can go to lower $t$ values but we lose cooling power
because of the $1/y$ factor and because of the dip in the curves$.$

Consider $d=10$ nm, $M=10^{11}$ which is perhaps within practical
reach.  The value $t=0.4$ ($T_{F}=35$ mK and $x=0.2\%)$ implies
$y=0.2$ and the cooling would be on the order of a 50 $\mu $W in this
ballistic-flow waveguide case.  Such a result would be quite a
remarkable cooling rate, but the assumption of perfectly smooth walls
is optimistic.  However, the inevitable coating of the pore walls by a
couple of layers of solid $^{4}$He will enhance the smoothness.

It perhaps seems more likely that the walls of the channel would cause
scattering, including back scattering, interband transitions, etc.
Thus we consider next a simple wall-limited diffusion or Knudsen flow
model. 
Our starting point is a kinetic equation for
the distribution function in collision-time
approximation\cite{Ziman},\cite{Rice}
\begin{equation}
\mathbf{v}_{p}\cdot \nabla n_{p}=-\frac{1}{\tau }\delta n_{p}
\end{equation}
where $\tau $ is the time between collisions with the wall. The left side
can be written as $-v_{pz}\partial n_{p}^{(0)}/\partial \epsilon _{p}%
\partial \mu /\partial z,$ where $n_{p}^{(0)}$ is the local equilibrium
distribution function and $\delta n_{p}$ is the correction to local
equilibrium. The gradient in chemical potential $\partial \mu /\partial z$
is proportional to the gradient in density. A much more rigorous approach to
such a kinetic equation is described in Ref. \cite{Meyerovich}. We consider
again the case of a very narrow channel containing banding. We solve for $%
\delta n_{p}$ and use it to compute, say, the particle flux as 
\begin{equation}
J_{N}=\frac{2 }{d^{2}h}\sum_{n}\int_{-\infty }^{+\infty }dp_{z}\frac{%
p_{z}}{m}\delta n_{p}  \label{JNdiff}
\end{equation}
From Eq.\ (\ref{dedt}) the cooling power is found to be 
\begin{equation}
C_{V}\frac{dT}{dt}=-\left( \frac{2A}{d^{2}h\beta ^{2}}\right) p_{K}
\end{equation}
where we have introduced the same prefactor as in the waveguide case so that 
\begin{equation}
p_{K}=d\frac{d\gamma }{dz}\left[ \frac{t}{y\left\langle \eta
_{n}\right\rangle }\right] ^{1/2}\sum_{n}\left[ G_{3/2}(\alpha
_{n})-\varsigma _{n}G_{1/2}(\alpha _{n})\right] .  \label{walllim}
\end{equation}
with $\zeta _{n}=\xi -y\eta _{n}/t.$ In this equation the collision time $%
\tau $ is writen in terms of the \emph{transverse} velocities in a channel.
That is, we write $\tau \sim d/(2\bar{v})$ with $\bar{v}=\sqrt{2\left\langle
\epsilon _{n}\right\rangle /m}$ where 
\begin{equation}
\left\langle \epsilon _{n}\right\rangle \equiv \sum_{n}\epsilon _{n}\int_{-%
\infty }^{+\infty }dp_{z}n_{p}^{(0)}(\epsilon _{n})\left[ \sum_{n}\int_{-%
\infty }^{+\infty }dp_{z}n_{p}^{(0)}\right] ^{-1}
\end{equation}
If we divide this quantity by the lowest band edge, then we have $%
\left\langle \epsilon _{n}\right\rangle /\epsilon _{0}=\left\langle \eta
_{n}\right\rangle $ used in Eq.\ (\ref{walllim}).

We have to evaluate the derivative $d\gamma /dz.$ What we mean by this
quantity is $d\gamma /dz=(d\gamma /dn)(dn/dz)$ since it is the gradient in
density $n$ that drives the flow. We can find this by taking the derivative
of the self-consistent expression, Eq.\ (\ref{selfconsis}). In that equation $%
t$ depends on $n$ because $\epsilon _{F}$ does. We find 
\begin{equation}
\frac{d\gamma }{dz}=\frac{2t}{3n}\frac{F_{1/2}}{tG_{1/2}}\frac{dn}{dz}%
\approx \frac{4}{9}\frac{G_{3/2}}{G_{1/2}}\frac{1}{L}=\frac{4}{9}\xi \frac{1%
}{L}
\end{equation}
where in the middle form we have taken the density gradient as $n/L$ with $L$
the length of a channel. What we compute then is 
\begin{equation}
p_{K}=\frac{4}{9}\frac{d}{L}\xi \left[ \frac{t}{y\left\langle \eta
_{n}\right\rangle }\right] ^{1/2}\sum_{n}\left[ G_{3/2}(\alpha
_{n})-\varsigma _{n}G_{1/2}(\alpha _{n})\right] .
\end{equation}

We again plot the product, $y~p_{K}$. The results are shown in Fig.~3; note
that the ratio $d/L$ is \emph{not} included in the plots. The curves differ
from those of the waveguide in a several ways. The curves are always
positive, i.e., represent cooling, and the $y=1$ degenerate case diverges
(the $T^{2}$ in the prefactor cures that). We have checked the numerical
calculations by doing various limiting relations analytically. For example,
the continuum (small $y)$ case in the limit $t\rightarrow 0$ is $%
p_{K}\rightarrow 2.41d/Ly$. 
\begin{figure}[h]
\centering
\includegraphics[width=3.0in]{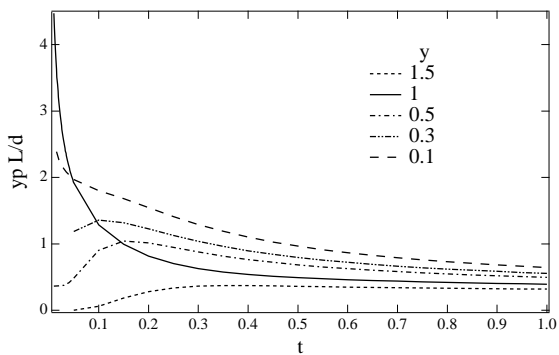}
\caption{Reduced Knudsen cooling power as a function of temperature
parameter $t$ for various $y$ values for a square cross section. The factor
of $d/L$ in the overall cooling power is not included here.}
\end{figure}

As in the waveguide case, small $y$ gives amplified cooling. Indeed, for $%
M=10^{11}$ holes, $y=0.1$ gives $p_{K}L/d\approx 1.6/y=16$. Again consider $%
d=10$ nm. The cooling power is $13$ $\mu $W$\times 16d/L=208d/L$ $~\mu $W. A
membrane width of $L=1$ $\mu $m gives a cooling of just $2.1$ $\mu $W. By
Eq.\ (\ref{tvsyd}) we have $t=0.18$ with the $y$ value chosen or $T_{F}=$ 83
mK and $x=0.5\%.$ This cooling power might be useful if the width $L$ used
is not too optimistic. The cooling would be enhanced if a larger $M$ value
were available. For small $y$ values the continuous degenerate limit gives
an upper limit on the cooling power within these conditions. We have 
\begin{equation}
p_{K}\lesssim 13\frac{2.4}{y}\frac{d}{L}\mathrm{\mu W}=\frac{31.2}{y}\frac{d%
}{L}\mathrm{\mu W}
\end{equation}

In summary, we have examined here the physics of particle flow through
narrow pores and estimated the possibility of cooling by this method
and have found some potential for success.  As nano-technology
improves, the possibilities may increase.  A very smooth-walled
channel that would provide waveguide type flow would give the greatest
cooling.  The more probable situation of Knudsen flow, while providing
cooling over all parameter ranges, has the factor $d/L$ reducing the
cooling power.  However, even that circumstance does not make it
impossible.  The numerical results give hope that this approach can
lead to a add-on device to extend the range of a dilution
refrigerator.  Experiments are being planned to test the potential of
the method.\cite{Owers} While we have considered the possibility of a
practical application of nanopores here, experiments on this kind also
provide interesting physics, namely detecting quantum size effects in
the narrow channels and the resultant restriction of states as already
seen experimentally.\cite{Nanophysics,Pickett}

We thank R. Hallock, J. Owers-Bradley, and G. Pickett for useful
conversations.

\end{document}